\documentclass[conference,compsoc]{IEEEtran}

\usepackage{cite}

\usepackage{url}
\usepackage[hidelinks]{hyperref}
\usepackage{amsmath,amssymb,amsfonts}
\usepackage{algorithmic}
\usepackage{graphicx}
\usepackage{textcomp}
\usepackage[dvipsnames]{xcolor}
\def\BibTeX{{\rm B\kern-.05em{\sc i\kern-.025em b}\kern-.08em
    T\kern-.1667em\lower.7ex\hbox{E}\kern-.125emX}}

\usepackage{tikz}
\usepackage{pgfplots}
\graphicspath{ {./images/} }

\definecolor{spdiff}{rgb}{0,0, 0}

\newcommand{\IV}{IV}
\newcommand{\salt}{sa}
\newcommand{\docid}{\textsf{docid}\xspace}
\newcommand{\dotdot}{..}

\makeatletter
\renewcommand{\paragraph}{%
  \@startsection{paragraph}{4}%
  {\z@}{1.2ex \@plus 1ex \@minus .2ex}{-1em}%
  {\normalfont\normalsize\bfseries}%
}
\makeatother

\newcommand{\smidge}{{\kern .05em}}

\newcommand{\getsr}{{\;{\leftarrow{\hspace*{-3pt}\raisebox{.75pt}{$\scriptscriptstyle\$$}}}\;}}

\newcommand{\calD}{\altmathcal{V}}

\newcommand{\advA}{\altmathcal{A}}

\newcommand{\bnm}{\begin{newmath}}
\newcommand{\enm}{\end{newmath}}
\newcommand{\bne}{\begin{newequation}}
\newcommand{\ene}{\end{newequation}}

\newenvironment{newmath}{\begin{displaymath}%
\setlength{\abovedisplayskip}{4pt}%
\setlength{\belowdisplayskip}{4pt}%
\setlength{\abovedisplayshortskip}{6pt}%
\setlength{\belowdisplayshortskip}{6pt} }{\end{displaymath}}

\newenvironment{newequation}{\begin{equation}%
\setlength{\abovedisplayskip}{4pt}%
\setlength{\belowdisplayskip}{4pt}%
\setlength{\abovedisplayshortskip}{6pt}%
\setlength{\belowdisplayshortskip}{6pt} }{\end{equation}}

\newcommand{\secref}[1]{Section~\ref{#1}}

\newcommand{\figref}[1]{Figure~\ref{#1}}

\newcommand{\verylongrightarrow}[1]             
      {\setlength{\unitlength}{.01in}           
      \begin{picture}(#1,1) \put(0,0){\vector(1,0){#1}} \end{picture}}

\newlength{\saveparindent}
\newlength{\saveparskip}
\newcounter{ctr}

\newenvironment{newenum}{%
\begin{list}{{\rm (\arabic{ctr})}\hfill}{\usecounter{ctr} \labelwidth=17pt%
\labelsep=5pt \leftmargin=22pt \topsep=3pt%
\setlength{\listparindent}{\saveparindent}%
\setlength{\parsep}{\saveparskip}%
\setlength{\itemsep}{2pt} }}{\end{list}}

\definecolor{clr3_1}{RGB}{203,106,73}
\definecolor{clr3_2}{RGB}{164,108,183}
\definecolor{clr3_3}{RGB}{122,164,87}

\definecolor{clr5_1}{RGB}{75,174,141}
\definecolor{clr5_2}{RGB}{202,86,136}
\definecolor{clr5_3}{RGB}{133,160,64}
\definecolor{clr5_4}{RGB}{135,116,202}
\definecolor{clr5_5}{RGB}{202,112,64}

\definecolor{clr2_1}{RGB}{179,102,158}
\definecolor{clr2_2}{RGB}{152,152,77}

\definecolor{clr6_1}{RGB}{167,221,226}
\definecolor{clr6_2}{RGB}{230,184,179}
\definecolor{clr6_3}{RGB}{155,194,175}
\definecolor{clr6_4}{RGB}{209,187,223}
\definecolor{clr6_5}{RGB}{212,217,182}
\definecolor{clr6_6}{RGB}{170,196,226}

\definecolor{clr6_1}{RGB}{121,113,168}
\definecolor{clr6_2}{RGB}{121,177,69}
\definecolor{clr6_3}{RGB}{154,72,190}
\definecolor{clr6_4}{RGB}{89,141,108}
\definecolor{clr6_5}{RGB}{183,73,89}
\definecolor{clr6_6}{RGB}{185,124,63}

\usepackage{xspace}
\usepackage{cryptocode}
\usepackage{amsfonts}
\usepackage{amsmath}
\usepackage{enumitem}
\usepackage{setspace}
\usepackage{adjustbox}
\usepackage{subfig}
\usepackage{booktabs}
\usepackage{siunitx}
\usepackage{tabularx}
\usepackage{xcolor}

\usepackage{physics}
\usepackage{amsmath}
\usepackage{tikz}
\usepackage{mathdots}
\usepackage{yhmath}
\usepackage{cancel}
\usepackage{color}
\usepackage{siunitx}
\usepackage{array}
\usepackage{multirow}
\usepackage{amssymb}
\usepackage{gensymb}
\usepackage{tabularx}
\usepackage{extarrows}
\usepackage{booktabs}
\usetikzlibrary{fadings}
\usetikzlibrary{patterns}
\usetikzlibrary{shadows.blur}
\usetikzlibrary{shapes}

\DeclareMathAlphabet{\altmathcal}{OMS}{cmsy}{m}{n}

\newcommand{\cell}{\mathsf{cell}}
\newcommand{\node}{\mathsf{node}}

\newcommand{\concat}{\:\|\:}
\newcommand{\calE}{\mathcal{E}}

\definecolor{lightgray}{gray}{0.9}
\definecolor{green}{rgb}{0,0.5,0}
\definecolor{purple}{rgb}{0.55, 0.0, 0.55}\definecolor{lightgray}{gray}{0.9}
\definecolor{green}{rgb}{0,0.5,0}
\definecolor{purple}{rgb}{0.55, 0.0, 0.55}
\definecolor{brown}{rgb}{0.59, 0.29, 0.0}

\newcommand{\deltamax}{\delta_\mathrm{max}}
\newcommand{\deltamin}{\delta_\mathrm{min}}
\newcommand{\thh}{\ensuremath{^\mathrm{th}}}

\begin{document}

\date{}

\title{Injection Attacks Against End-to-End Encrypted Applications}

\author{
  {\rm Andrés Fábrega$^{1}$, Carolina Ortega Pérez$^{1}$, Armin Namavari$^{1}$, Ben Nassi$^{2}$, Rachit Agarwal$^{1}$, Thomas Ristenpart$^{1, 2}$}\\
  $^{1}$ Cornell University \hspace*{3em} $^{2}$ Cornell Tech
} 

\maketitle
\thispagestyle{plain}
\pagestyle{plain}
\begin{abstract}
%
%
We explore an emerging threat model for end-to-end (E2E) encrypted applications:
an adversary sends chosen messages to a target client, thereby ``injecting''
adversarial content into the application state. Such state is subsequently
encrypted and synchronized to an adversarially-visible storage. By observing the
lengths of the resulting cloud-stored ciphertexts, the attacker backs out
confidential information. 

We investigate this injection threat model in the context
of state-of-the-art encrypted messaging applications that support E2E encrypted
backups. We show proof-of-concept attacks that can recover
information about E2E encrypted messages or attachments sent via WhatsApp,
assuming the ability to compromise the target user's Google or Apple account
(which gives access to encrypted backups).
We also show weaknesses in Signal's
encrypted backup design that would allow injection attacks to infer metadata 
including a target user's number of contacts and conversations, should the
adversary somehow obtain access to the user's encrypted Signal backup.

While we do not believe our results should be of immediate concern for users of
these messaging applications, our results do suggest that more work is needed to
build tools that enjoy strong E2E security guarantees.

\end{abstract}

\section{Introduction}
Deployment of end-to-end (E2E) encryption has improved the confidentiality and
the integrity of data in various contexts, including messaging~\cite{whatsappstats,
messengerstats,signalstats}, cloud storage~\cite{icloudsecurity, dropboxsecurity}, and other web
applications~\cite{googledocssecurity}.  The security of E2E encrypted
messaging protocols~\cite{garman2016dancing,patersonthree,tyagi2019traceback,
rosler2018more, cohn2020formal, bogos2023security,sutikno2016whatsapp,
endeley2018end, canetti2022universally, herzberg2020secure, alwen2019double,
jost2019efficient, bienstock2022more,
albrecht2022practically} and file storage~\cite{backendal2023mega, harnik2010side,yang2020data}
has been studied extensively, giving us confidence that even sophisticated,
nation-state level adversaries cannot violate the security of 
state-of-the-art E2E encryption tools without compromising endpoint devices.

To support new features, the complexity of E2E encryption tools is increasing.
Messaging applications have recently started to provide backup features that
allow users to recover their messages when they need to transition to a new
device. \mbox{WhatsApp}~\cite{whatsappe2ebackups1,whatsappe2ebackups3} and
Signal~\cite{signalencryptedbackups}, which together account for billions of
users~\cite{whatsappstats, signalstats}, both have
opt-in backup features. WhatsApp provides
automatic upload of backups to a user's Google Drive or iCloud accounts, while
Signal allows users to manually export them.  
In both cases, backups are
encrypted, and should only be decryptable by the legitimate user~\cite{whatsappe2ebackups1}. 
Therefore backups should enjoy the same level of confidentiality as E2E encrypted messaging.

In this work, we introduce new attacks against E2E encrypted messaging
applications.
Our most damaging attacks recover partial information about messages or
attachments sent from one honest user to a target
honest user~$U$. The attacker needs the ability to send adversarial
messages to the target user—--thereby ``injecting'' adversarial
content into the application state—--and the ability to observe
the target user's encrypted backups.
As such, we refer to these as injection
attacks. Our attacks do not invalidate the security of the E2E encrypted 
messaging protocol used (the Signal protocol in both cases), 
but rather violate confidentiality via cryptographic vulnerabilities in other parts
of the application, namely, their backups.

We stress that, in our threat model, the attacker never has access to the backup
decryption key. Thus, a priori, an adversary should not be able to learn
about~$U$'s conversations with other honest users. To see if this holds true in
practice, we perform a security analysis of both WhatsApp and Signal in the
context of injection attacks. 

\begin{figure*}
    \centering
    \footnotesize
    \renewcommand{\arraystretch}{1.1}
    {\color{spdiff}
    \footnotesize
      \begin{tabular}{l l l l r r} 
     \toprule
      \textbf{Application} & \textbf{Attack} & \textbf{Attack Vector} & \textbf{Setting} & \textbf{Backups seen} & \textbf{Messages sent}  \\
     \midrule
      \multirow{3}{*}{WhatsApp} & (1) Dictionary attack on attachments & Attachment deduplication & Noisy
      device & $q$ & $\sum_{i \in [q-1]} \lceil n/16^i \rceil $ \\[1pt]
      & (2) Dictionary attack on messages & zlib compression & Quiet device & $2 \cdot \lceil \log_2 n \rceil + 1$ & $2 \cdot \lceil \log_2 n \rceil + 1$\\[1pt] 
      & (3) Distinguishing attack on messages & FTS4 index & Quiet device & 16 & 44 \\[3pt]
      Signal & (4) Learn number of contacts, messages & Serialization method & Noisy device & 1 & 1 \\
     \bottomrule
\end{tabular}}
\caption{Summary of the attacks discovered in this work. Here $n$ is the size of the dictionary $\calD$ (possible attachments or messages) and $q = \lceil \log_{16} n\rceil$. }
\vspace{-0.3cm}
\label{fig:attacks-summary}
\end{figure*}


For WhatsApp, we identify three distinct attack vectors. First, attachments such as
images, videos, or PDF files are deduplicated---the backup only stores
one encrypted copy of each unique attachment---even if it was received from different
senders. Second, the serialized database file is compressed using a
standard library (zlib) before it is encrypted. Third, WhatsApp uses an SQLite
module (FTS4) to build a search index for all messages across all
conversations. In all three cases, the 
\emph{length} of the encrypted backup ends up being a function of content from
different senders, and serves as a side-channel through which
the adversary can deduce information about honest messages.

In some special cases, exploitation is straightforward for the
adversary: for media deduplication, the adversary can observe a backup, send $U$ a candidate media file, and observe
the subsequent backup. If $U$ has already received this file
before, the size of the second backup ciphertext will grow by
less than what is expected. We show a more sophisticated attack approach that is
robust to most kinds of noise (other, unrelated activity on $U$'s device such as receiving
other attachments beyond the target messages) and allows determining which of
$n$ attachments were received by $U$. It requires the adversary to observe 
at most $\lceil \log_{16} n \rceil$ backups.
We note that deduplication exploits
have been considered in other contexts, such as encrypted
storage~\cite{harnik2010side}; however, as far as we know, this is the first
work that shows this issue arises in encrypted messaging backups.

Exploiting zlib compression required understanding complex interactions between
database serialization and compression, and how sending messages affects the
resultant backup. Nevertheless, we show a binary-search style injection attack
that determines which of~$n$ messages was recently received by~$U$, by
adaptively injecting at most $2 \cdot \lceil \log_2 n \rceil + 1$ messages and observing the same
number of backups. We have demonstrated this attack 
in a lab setting where the victim $U$'s client application has no other 
activity during the attack.
We call this scenario the ``quiet device'' setting. We
experimentally verify the attack for small~$n$.

Finally, we show that even if zlib compression and deduplication were turned off, 
there are additional sources of leakage that stem from the use of a text keyword search index
called FTS4.  
This is a delicate vulnerability that arises due to subtle interactions between WhatsApp's
use of FTS4, the inner mechanics of the B-tree data structures used to store the
index, and SQLite serialization.  Our attack consequently is technically
complicated, in large part because the adversary does not initially know the
internal state of the target's data structures, and must account for this by
adaptively modifying the state via injections to enable learning confidential
information.  Nevertheless, we experimentally demonstrate that in the quiet device
setting, an attacker can determine which of two messages~$U$ has received from an honest party
by injecting at most 44 messages and observing at most 16 backups. This attack
vector may be of relevance to other applications that index mixtures of trusted
and untrusted data using FTS4 (e.g., multi-tenant search services~\cite{wang2017side}).

We also explored Signal, whose bespoke serialization and encryption mechanism
avoids many of the problems above. Nevertheless, we built an attack which
exploits the fact that the structure of their backups leaks the size of each
row in the target $U$'s client-side database. Combining an injection attack with
additional heuristics, an adversary can infer which rows are part of which
tables, by observing only two backups.  This allows them to learn the number of
messages~$U$ has received,~$U$'s number of contacts, and more.  We
discuss how this attack can also be adapted to work in the ``noisy
device'' setting, either assuming the size of messages sent
by honest parties (the noise) have bounded length, or by injecting a small
sequence of random-length messages. These only require a single backup.

\paragraph{Contributions} We are the first to explore injection attacks against state-of-the-art
encrypted messaging applications that utilize encrypted backups. 
{\color{spdiff}A summary of our attacks appears in 
Figure~\ref*{fig:attacks-summary}.} These
demonstrate how to violate the confidentiality of messages and attachments sent
on WhatsApp and, for Signal, how to efficiently reveal potentially sensitive
metadata about a user's contacts and messages. While our in-lab experiments
do not indicate that injection attacks are an immediate threat to user privacy,  
they do highlight previously unrecognized challenges faced when attempting to
achieve E2E confidentiality guarantees in adversarial settings.

We therefore 
discuss potential countermeasures in the body. While some attacks have 
straightforward mitigations, others uncover the need for additional work to
find solutions to building backups that are both efficient and secure.
Beyond backups, and given the expanding set of applications being built with E2E
encryption guarantees in mind, our results motivate future work on principled
mechanisms for discovering and mitigating injection attacks.

\paragraph{Ethics and responsible disclosure} Our experiments involved researcher accounts that were not
used for other purposes and minimal load on Signal and WhatsApp
services, requiring a small number of messages sent at a reasonable pace. To
see how attacks can scale up in a way that might load servers, we implemented simulators
whose results we validated via smaller manual experiments with real
clients.

We disclosed our findings to Signal and WhatsApp. Signal acknowledged the vulnerabilities and deployed mitigations in the v1 revision of their Android backup file format,\footnote{Mitigations rolled out in Android builds following commit c6473ca9e63236af3eae9959a50cfa643d53272e in their open-source repository~\cite{githubsignalandroid}.} per our recommendations from Section~\ref*{sec:discussion}. WhatsApp acknowledged receipt of our disclosure, and awarded us a bug bounty. At the time of writing, they have not yet deployed mitigations.
\section{Related Work}

\paragraph*{Analysis of E2E encrypted applications} Recent work has explored attacks on a
variety of E2E encrypted platforms. Paterson et al.~\cite{patersonthree} present seven
attacks on Threema, an E2E encrypted messaging platform, across three different threat
models. One of their attacks targets encrypted backups; we discuss this further below. 
Albrecht et al.~\cite{albrecht2022practically} provide novel attacks on
Matrix, a federated E2E encrypted messaging platform and open protocol. Backendal et
al.~\cite{backendal2023mega} demonstrate attacks on MEGA, an E2E encrypted file-sharing
service. None of these attacks work in our threat model, requiring adversarial
capabilities we do not assume. They also exploit vulnerabilities in the (weak) 
cryptographic schemes and 
protocols used, while ours do not. 
That said, these recent prior results and ours together showcase a growing
need to more holistically evaluate the security of E2E encrypted applications.

Despite this, and even though WhatsApp and Signal have
been the source of much academic work~\cite{rosler2018more, cohn2020formal,
bogos2023security,sutikno2016whatsapp, endeley2018end, canetti2022universally,
herzberg2020secure}, little attention has been placed on their backups. 
A security assessment of WhatsApp's backups scheme by NCC
Group~\cite{nccwhatsapp} explicitly mentions that the ``backup encryption
implementation'' was not in scope, and they focused solely on key management and
privilege separation. There is no overlap between our findings and their report. More importantly, a recent paper~\cite{davies2023security} provides an extensive formal security analysis of WhatsApp's backup system under the universal
composability (UC) framework, and found that the protocol ``provides strong protection of users’ chat history and passwords''. However, their analysis was limited to the ``password-protected
key retrieval (PPKR) scheme'' of WhatsApp's backups system, whereas our attacks focus on details of the encryption mechanism itself.

\paragraph*{Length-revealing encryption} Many works have sought to exploit the
lengths of ciphertexts as a channel for learning confidential information.
Perhaps most relevant is that our attacks bear some similarity to the literature
on traffic analysis (c.f.,~\cite{dyer2012peek}) that has been explored against
network security protocols like TLS and SSH. These use the lengths of
ciphertexts to infer information about plaintext communications, such as the
website being visited over an encrypted tunnel
(e.g.,~\cite{panchenko2016website,hayes2016k}). But these attacks are 
tailored to the network setting and do not work in our context.

\paragraph*{Attacks on compression before encryption} It has long been
recognized that compression (such as zlib) before encryption can result in
vulnerabilities~\cite{kelsey2002compression}. For example, attacks have been given against
TLS when using
compression~\cite{rizzo2012crime,gluck2013breach,vanhoef2016heist}, the iMessage
E2E encrypted messaging protocol~\cite{garman2016dancing},  and the aforementioned
Threema attack~\cite{patersonthree}. The iMessage attack exploits compression
of chat messages before encryption; this is now known to be bad practice and
state-of-the-art E2E encrypted messaging systems do not use message compression. 

Threema is a messaging application advertising E2E encrypted messaging, but Paterson et
al.~\cite{patersonthree} showed a number of flaws in its cryptographic protocols.  One of their
attacks targets the fact that Threema supports encrypted backups. Their attack,
like ours from \secref{sec:compression-attacks}, uses an injection-type
attack exploiting zlib compression and a
length side-channel. But their attack requires physical access to the target's
handset and the ability to unlock it. Even if one assumes physical access to a
target device, their attack doesn't apply to WhatsApp or Signal because it
exploits details of Threema's design, in particular the ability to target
recovery of a secret key stored within the backup. 

Concurrent work introduced the DBREACH attack~\cite{hogan2023dbreach}, which explores compression 
specifically in the context of databases. 
However, it operates in a different threat model, for two main reasons: (1) their attacks assume an adversary who can insert \emph{and} edit content in the database, whereas WhatsApp does not allow users to edit sent messages; (2) more importantly, their attacks assume that the adversary controls the entire payload of the insertions they trigger (e.g., by being able to send direct SQL statements to the database file). Sending a message in WhatsApp, however, is a much more noisy interface: it inserts a lot of information outside of the adversary's control into the database, such as timestamps and other metadata. As such, their attacks are not applicable to this setting.

Another form of compression is deduplication, in which systems only store one copy
of duplicate files. Like zlib-style compression, deduplication before encryption
has been explored as a vulnerability in other contexts such as client-side
encrypted file storage~\cite{harnik2010side,halevi2011proofs,bosman2016dedup}.
Our injection attacks exploiting deduplication
(\secref{sec:dedup}) are similar to these prior attacks, but require additional refinements due to details of WhatsApp's architecture. Our results
highlight how deduplication vulnerabilities arise in new settings.

\paragraph{Attacks on encrypted search and databases} A now long line of work
exists on leakage-abuse attacks against encrypted databases and search
indexes~\cite{cash2015leakage,blackstone2020revisiting,giraud2017practical,grubbs2016breaking,islam2012access,naveed2015inference,pouliot2016the,rompay2017a,wright2017early,zhang2016all}. These use passive observations of
accesses to an encrypted data store by an adversarial cloud service to 
violate confidentiality. These do not apply to our setting, where access leaks
only the fact that a backup is being stored or downloaded. 

In addition to passive attacks, this literature explored active injection
attacks~\cite{cash2015leakage,zhang2016all,xu2021searching} that arise in a
threat model similar to ours.  Prior injection attacks
including~\cite{cash2015leakage,zhang2016all} take advantage of leakage of
search patterns (what documents match against a keyword search) or access
patterns (which documents are accessed due to a search) to recover plaintext
content, and do not apply to our setting where search does not involve
interactions with the adversarial storage service.  Our attacks are closer to a
vulnerability highlighted by Xu et al.~\cite{xu2021searching} that showed how
the length of encrypted Lucene search indexes might be exploitable for plaintext
recovery via an injection attack. But they stop short of giving a fully
specified attack, their results do not apply to our setting (which does not
involve Lucene), and it's unclear whether their attack affects any deployed
system.

\section{E2E Encrypted Backups and Threat Models}
\label{sec:architecture}

In this section, we provide a broad overview of the architecture of end-to-end
(E2E) encrypted backups, and detail the threat models that we explore.

\paragraph{Basic messaging architecture} For our purposes, a messaging service
consists of a client~$U$ running on some user device. The client sends and
receives E2E
encrypted messages using some service-operated servers. The 
client stores relevant information in a
local database~$D$, stored inside the device's internal storage.
We denote by $V$ the set of possible inputs to the database.

If the service supports E2E encrypted backups, periodically, a snapshot $S$ of the
database $D$ is generated, (optionally) compressed, and then encrypted.  The
encryption uses some application-specific symmetric encryption scheme with a
secret key~$K$.   The resulting ciphertext~$C$ is stored either locally and/or
in some cloud storage provider, such as Google Drive or iCloud. We refer to the
location where backup ciphertexts are stored as the backup storage. Since
backups can be periodic, we use $S_1,\ldots,S_q$ to denote a sequence of
generated snapshots, and $C_1,\ldots,C_q$ to denote the associated ciphertexts
sent to the backup storage. 

What makes the backups ``end-to-end'' encrypted is that only the user and their
device know the key~$K$ required to decrypt the backups; no one else---including
the messaging service provider nor a storage provider for backups---should know
the secret key.

\paragraph{Example: WhatsApp backups} In late 2021, WhatsApp began providing
opt-in E2E encrypted backups for their users~\cite{whatsappe2ebackups1}. 
It supports two types of secret key~$K$: a 64-digit secret chosen uniformly at
random by the device, or a short password from which $K$ is derived using an
oblivious pseudorandom function (OPRF)~\cite{jarecki2018opaque} whose secret is only stored within
hardware security modules managed by WhatsApp~\cite{whatsappe2ebackups1, whatsappe2ebackups2,
whatsappe2ebackups3}. In the latter case, recovering
$K$ would require not only knowing the target's password, but also control over
their phone number. Our attacks assume no knowledge of $K$ and are 
agnostic to how~$K$ is generated. 

There is less documentation about other aspects of how backups are generated. We
reverse engineered WhatsApp version 23.2.75 for Android.
WhatsApp stores the data for a user on an SQLite3 database sitting in their
device's storage---which represents $D$. 
The schema for this database is quite complex, involving over 100 tables; we
highlight particularly relevant ones later.
WhatsApp serializes $D$ using SQLite's
standard database file format~\cite{sqlitedatabasefileformat}. We provide more details 
on this format in \secref{sec:compression-attacks}. This is then compressed using zlib (using compression level 1).\footnote{zlib compression levels range from 0
to 9, and specify trade-offs between speed and compression.} The output of zlib
is then encrypted using AES-GCM~\cite{mcgrew2004galois} to yield the encrypted
backup~$C$, which is then uploaded to the cloud. In fact, $C$ has slightly more structure with extra checksums that
are unimportant for our results.

In addition to the monolithic encryption of the database, WhatsApp stores in the cloud an encryption of each individual attachment (images, videos, PDFs, and any arbitrary file type) sent to and received by the user. These are encrypted using a symmetric encryption algorithm, the low-level details of which are opaque to us (e.g., the cipher used and how attachment keys are managed); however, these are not relevant to our attacks. Alongside each attachment ciphertext, WhatsApp stores a series of plaintext headers containing metadata such as the time the ciphertext got uploaded, its size, etc. As we will see in Section~\ref{sec:dedup}, these headers will be crucial to our attacks.

WhatsApp backups are opt-in, though it periodically encourages users to
turn it on with a pop-up. When configuring it, users can specify that backups
should occur manually, or automatically with daily, weekly, or monthly
frequency at a fixed time (always at 2:00 am). On Android devices, backups are stored on  Google Drive
under the user's account.

\paragraph{Example: Signal backups} We also investigated Signal, using v6.22.8
for Android. Signal similarly stores~$U$'s data in an SQLite3 database,
which represents $D$. The schema of the database includes 42 tables. The database
$D$ is serialized by generating a sequence of SQL commands that suffice to allow
reconstructing, from a fresh SQL database instance, the state of $D$. Why Signal
takes this approach rather than WhatsApp's is unclear. Each SQL command is
separately encrypted using AES in counter mode with HMAC to generate an
authentication tag, with a plaintext length header prepended to the resulting
ciphertext. Media such as images, videos, or other sent files is handled by
encrypting a header indicating the length of the media, and then concatenating
to that an encryption of the media contents. More details are given in
\secref{sec:passive-attack}. The key used for encryption is derived from a
randomly generated 30-digit passphrase (the user must write it down or otherwise
store it elsewhere).

Once enabled, backups are created by Signal automatically once a day. 
The user can also manually trigger a backup. 
Signal does not have built-in support for storing backups
at a cloud provider. Backups can only be created on
Android devices, and stored locally --- documentation suggests that users manually
copy the backup to another device~\cite{signal-backup}. 
However, if users want to
prevent losing their data, they can
arrange for the local backup to be synchronized as a file to an 
external cloud storage service using any one of many possible cloud storage tools.
This is not recommended by Signal; however,
there are third-party tutorials online to synchronize Android
folders with cloud storage services~\cite{androiddrivesync,signaldrivesync}.

\paragraph{Threat models}\label{sec:threat-models} 
The backup systems are designed to provide E2E confidentiality, even in the face
of sophisticated adversaries. Both applications derive secure keys, barring simple
approaches like brute-force password cracking attacks. 
Instead, we focus on a subtler threat model that gives rise to what we call
\emph{injection attacks}.

Our threat model assumes that the adversary can interact with a target
client~$U$ via the messaging application, e.g., to send it one or more messages
using the standard E2E encrypted channels. The adversary just uses the normal
interface for sending messages, and can even do this using unmodified client
software. 
The adversary can then observe the next backup ciphertext. 
This process of sending messages and observing ciphertexts can be repeated,
allowing the adversary to adaptively select subsequent messages to send as a
function of previous backup ciphertext observations.   

In more detail, consider a target client~$U$ that has some initial database
$D_0$.  The adversary observes the backup $C_0$ associated with $D_0$. The
adversary controls one or more adversarial clients that can now interact
with~$U$ in any way allowed normally by the messaging protocol; in the case
studies that follow, this includes sending regular chat messages or attachments to~$U$. 
This results in a new
state $D_1$, which therefore contains some \emph{injected} adversarial content.  
The adversary waits for~$U$ to generate a new backup~$C_1$ associated to $D_1$ and
observes it.  The adversary can repeat this process, sending more commands to
generate a new state~$D_2$, observe $C_2$, and so on. Later we will use~$q$ to denote the
number of rounds of backup observation.

The adversary's goal is to learn some confidential information. For example, the
adversary may attempt to infer whether~$U$ had previously received a message $m$
falling in some adversarially-known set of possibilities~$\calD$. If $|\calD| =
1$, we call this a confirmation attack; if $|\calD| = 2$, a distinguishing
attack; and if $|\calD| > 2$, a (partial) message recovery attack.  
Message contents are just one type of adversarial goal, and we will also explore
injection attacks that reveal other information about $U$, including metadata
such as their number of conversations or contacts.

\paragraph{Discussion} We turn to the practical relevance of our injection threat model.
Five key assumptions underlie our attacks: (1)~access to backup
storage; (2)~users ignoring injected messages; 
(3)~reasonably frequent backups; (4) some of our attacks rely on the target device being ``quiet''; and (5) the victim has not removed the target message from their device.

The first assumption is standard, and may arise from an insider attack or
compromise of the cloud storage provider itself, or just from access to the
user's cloud storage account (e.g., for WhatsApp, having the ability to log into
someone's Google Drive or iCloud account---for Android and iOS,
respectively---suffices to retrieve backup ciphertexts). Signal suggests users
store backups on a separate device and does not natively support cloud backups,
but realistically some users will want the durability of the cloud.
Online help articles by third parties not associated with Signal (e.g.,~\cite{androiddrivesync,signaldrivesync}) provide instructions for
configuring Android folders to automatically synchronize to a cloud storage
service.

We emphasize that compromising a target's cloud storage account does \emph{not}
give the adversary the ability to take over the target's E2E encrypted messaging account.
Messaging takeover requires control over the phone number (e.g., via a SIM
swapping attack) and then porting the account to a new, adversarially
controlled device. This is visible to the target, who would lose access to their
phone number. It also would not violate the confidentiality of past
communications (which need to be restored via the encrypted backup). 

Our injection attacks will instead allow ongoing, covert monitoring of a target
device, with exception being that the injected messages will be visible to a
user. We conjecture that many users would just assume these to be spam from
unrecognized numbers, and indeed our attacks can be extended in most cases to
allow some extra content to make the messages seem innocuous. A user may block
an unrecognized number, but the attacker can use different numbers to send
messages. In any case, we do not want security to rely on user vigilance. 

The frequency of backups may be limiting to attackers.  In both WhatsApp and
Signal, backups can be set to be automatic, with the default in both cases being
once per day. As far as we are aware, there is no way an attacker could trigger
a backup remotely.  Even so, a patient attacker can nevertheless
just wait for backups to occur.

 The level of activity on the target device impacts some of our attacks. In the ``quiet device'' setting, we assume that~$U$ is not interacting with other, non-adversarially controlled clients while the attack proceeds. This is realistic for users who only use their messaging client sparingly, but not for others who use it more frequently. In any case, security guarantees should not depend on a user being active or not. On the other hand, in the ``noisy device'' setting, we assume that the victim can interact with non-adversarial clients while the attack is running. This is a more practical context, as arbitrary users are vulnerable irrespective of their activity patterns.

The last key assumption of our attacks is that the victim does not remove the target message from their phone after it is received, i.e., that the message is not unsent, deleted, or auto-dissapears.\footnote{WhatsApp and Signal both have features for auto-dissapearing messages~\cite{whatsappautodisappear,signalautodisappear}.} If the user does so before the attack starts, this is essentially equivalent (for the purposes of our attack) to not receiving the message to begin with.

\section{Injection Attacks against WhatsApp}
\label{sec:injection}
Recall that in our injection attack setting, the adversary can interact with the
target client in between backups and observe the resulting ciphertexts.
The sizes of such ciphertexts are a side-channel by which confidential
information can be gleaned. Underlying our attacks against WhatsApp backups are three issues:
deduplication of attachments, compression of backups before
encryption, and subtle ways in which internal data structures of WhatsApp's database maintain state. While the first two issues have been explored in  other contexts before~\cite{harnik2010side,gluck2013breach, rizzo2012crime, kelsey2002compression, vanhoef2016heist}, our attack setting differs. To the best
of our knowledge, no prior work has explored the third source of side-channel
leakage.

\subsection{Exploiting Deduplication}\label{sec:dedup}

Deduplication-based attacks have been explored in the context of encrypted
storage~\cite{harnik2010side}. Here we demonstrate how to exploit deduplication in the context of E2E encrypted applications, namely, the E2E encrypted backups
as implemented by WhatsApp. To do so, we devise a file-recovery attack on the attachments received by the victim.

\paragraph{Attack vector}
Our close study of WhatsApp's architecture revealed that they perform attachment  deduplication---that is, store only one copy of an attachment in the backup---for many types of attachments, including images, videos, and PDF files. To do so, the client compares the SHA256 checksum of each new attachment against all prior ones, and stores it only if there is no match. Importantly, WhatsApp performs deduplication even if the attachment was received multiple times from \emph{different} senders. As we will see, such a deduplication mechanism enables cross-user attacks under our injection attack threat model.

\paragraph{Background}
Before explaining our full attack, we provide some additional context and notation regarding encrypted attachments.

Let $A \coloneqq \{(a_1, f_1), ..., (a_m, f_m)\}$ be pairs of attachments and their associated filenames, which the victim has already received. For now, we assume all attachments are distinct for clarity of presentation. When a user receives an attachment, their WhatsApp client stores the file with the same name as it was received. That is, the filename of the saved file is under the sender's control. The recipient can change the filename directly in the device's file system. However, the adversary can exploit the predictable periodicity of the backups to prevent the victim from having time to change the names, by performing injections shortly before a backup occurs. Importantly, Android and iOS both have a 255-character limit for filenames, which is important for our~attack.

To create a backup, each new attachment and filename\footnote{In fact, there is other metadata encrypted with the filename, but this is irrelevant to our attack, as its length remains constant across attachments.} gets encrypted separately with symmetric encryption schemes; old attachments that are already in the backup are not re-uploaded. Each attachment $a_i$ is encrypted into a ciphertext $c_i$, where $|c_i| = |a_i| + 16$, due to a 16-byte authentication tag. As such, there is a one-to-one correspondence between attachment ciphertext sizes and plaintext sizes. Each filename $f_i$ is encrypted using a block cipher in some mode of operation (presumably CBC) into a ciphertext $c'_i$. Therefore, the size of filename ciphertexts always increases in 16-byte jumps. The ciphertexts $C \coloneqq \{(c_1, c'_1), ..., (c_m, c'_m)\}$ are included in the victim's backup alongside the encryption of the main database file, where each $c'_i$ is stored as metadata for the associated $c_i$.

An adversary with access to the victim's backup can see the set of pairs $C$. Then, sending a new batch of $n$ attachments adds (at most) $n$ new ciphertext pairs to the backup. However, the order in which a new batch of attachments is uploaded between every two backups follows no discernible pattern, as far as we can tell. So, even though the adversary can identify which ciphertext pairs are new (by looking at the pairs present in the second backup but not in the first), they cannot directly map each new pair of ciphertexts to the corresponding attachments they sent.

\paragraph{File recovery attack}
We now describe how to leverage the deduplication mechanism to build a file recovery attack. Given a set of candidate attachments $\calD \coloneqq (v_1, ..., v_n)$, the adversary wants to determine which $v_i \in \calD$, if any, is present in the user's device. Our attack requires observing at most $\lceil \log_{16} n \rceil$ backups. Further, as we discuss later on, this attack works in the noisy device setting.

We first note that if all attachments in $\calD$ are of different sizes, there is a straightforward attack: the adversary can send all files in $\calD$ to the victim between two backups, and compare the sizes of the (at most) $n-1$ new attachment ciphertexts $c_{m+1}, ..., c_{m + n-1}$ in the cloud against the sizes of the files in $\calD$. As explained earlier, even though the adversary can identify which ciphertexts are new, these ciphertexts appear in arbitrary order. Even then, however, if there is no ciphertext of length $|v_j| + 16$, then attachment $v_j$ got deduplicated. (Filename ciphertexts are not relevant in this simplified setup.) This attack would require only two backups, regardless of the size of $\calD$, and reveals all members of $\calD$ received by the victim.

The attack above, albeit simple, may fail if there are repeated sizes in $\calD$. If there are two attachments $v_i$ and $v_j$ such that $|v_i| = |v_j|$, but there is only one ciphertext of size $|v_i| + 16$, the adversary has no way to tell which of these attachments corresponds to the ``missing'' ciphertext, and which corresponds to the uploaded ciphertext.

The key idea behind our attack is to use filenames as a means to differentiate the attachment ciphertexts, and thus relate them to the plaintext files. Since filenames represent a second degree of freedom through which the adversary can inject information, they can pick filenames such that every \emph{pair} $(c_i, c'_i)$ is of a unique length, instead every file being of a unique size. We now describe the full attack.

\begin{newenum}
    \item \emph{Pick filenames:} Evenly partition the list of candidate attachments $\calD$ into 16 classes $F_1, F_{17} ..., F_{241}$, where $F_i$ contains $\lceil n/16 \rceil$ files if $i \leq n$ mod 16, and $\lfloor n/16 \rfloor$ otherwise. For each file in $F_i$, pick an arbitrary filename of length $i$, and rename the file to this.
    
    As mentioned earlier, since filenames are encrypted using a block cipher, their lengths need to be picked in 16 byte jumps to trigger new block allocations. The limit of 255 characters, however, requires us to use a recursive strategy, as the adversary can only use $\lfloor 255/16 \rfloor + 1 = 16$ distinct lengths.
    
    \item \emph{Inject:} Send all $n$ candidate files to the victim.
    
    \item \emph{Backup:} Wait for the next backup, which now includes at most $n-1$ new pairs of ciphertexts $\{(c_{m+1}, c'_{m+1}), ..., (c_{m+n-1}, c'_{m+n-1})\}$.
    
    \item \emph{Find the missing size:} Inspect all $c'_{m+1}, ..., c'_{m+n-1}$ to identify which filename size $s$ was not uploaded to the backup: there will be $|F_s| - 1$ values for the filename ciphertext length of the deduplicated file. This implies that the file that got deduplicated was one of the files in $F_s$, although it is not clear which. 
        
    \item \emph{Recurse:} If $|F_s| > 1$, repeat steps 1--4 on (only) the files within $F_s$ as the input list of candidate files, i.e., $\calD = F_s$. Otherwise, output $F_s$ as the target attachment.
    
    Before proceeding to the next the recursive call, the adversary needs to \emph{unsend} all attachments sent in this round; otherwise, all files will already be present in the backup, so they will all get deduplicated, instead of just the one matching the target file. Importantly, unsending an attachment indeed removes it from the recipient's backup, so subsequent phases of the attack can still run.
\end{newenum}

This attack finds the target file after observing just $q \coloneqq \lceil \log_{16} n \rceil$ backups and sending at most $\sum_{i \in [q - 1]}(\lceil \frac{n}{16^{i-1}} \rceil)$ messages, because at most $\lceil \frac{n}{16^{i-1}} \rceil$ attachments are sent in the $i_{\textsuperscript{th}}$ recursive call. For example, it takes at most three backups (i.e., three days) and 4266 messages to find the target file within a list of 4000 candidate files.

Our attack can be extended (at the cost of more backups) to find \emph{all} matching files, instead of just one. In this case, there will be multiple values of $s$ in step (4) for which there are less than $|F_s| - 1$ filename ciphertexts, so the adversary can independently recurse into all $F_s$.

\paragraph{Assumptions} Our attack assumes that the attachments sent by the adversary are downloaded upon receiving them, either automatically by the client or manually by the user, as otherwise they are not incorporated into the backup. Automatic downloads from attachments received from contacts is the default setting in WhatsApp~\cite{whatsappautodownload}. For non-contacts, there are other potential roundabout ways in which the client will still automatically download a file, such as sending the attachments to a group chat which both the victim and adversary belong to.

In addition, our attack places only one constraint on external noise: it requires that, in every recursive call, none of the received attachments have the same file size \emph{and} the same filename size as the deduplicated attachment. More concretely, let $\hat{\calD} \subset \calD$ and $E$ be the injected attachments and the external attachments received in the same recursive call, respectively. Then, our attack assumes that, for every $(v, f) \in \hat{\calD}$ that gets deduplicated, it follows that, for all $(v', f') \in E$, $|v| \neq |v'| \wedge |f| \neq |f'|$. Otherwise, if one of the external attachments has the same size as the candidate file that gets deduplicated, the adversary will see a new ciphertext pair of the expected size, without being able to tell that this is a false negative. Our attack is robust to other types of activity on the target: the victim can receive arbitrary text messages and attachments (of other sizes) at any moment, and the attack still succeeds.

\paragraph{Implementation} We experimented with PNG images, MP4 videos, PDFs, and Word documents, but we believe the attack works for all attachment types. For each file type, we ran a proof-of-concept of the attack using 30 candidate files (all of the same size and with arbitrary content). We picked one of the 30 files at random and sent this to the victim from some non-adversarial device, to serve as the target file. We then ran the attack from an adversarial device. Additionally, we simulated external noise by sending 10 text messages and 10 attachments (of the same type, but of sizes outside of those in the candidate set) from the non-adversarial device to the victim device between every two backups. Our attack was able to successfully retrieve the target file in all cases.

\subsection{Exploiting Compression}\label{sec:compression-attacks}
We now turn to injection attacks on WhatsApp that exploit another mechanism: use of zlib to compress the serialized database file before encrypting it. As we will see, exploiting compression in the context of messaging applications with E2E encryption comes with various challenges, due to the ``noisy'' and limited injection interface (sending messages) and the complex ways in which the contents of messages are arranged in the serialized database file.

\paragraph{Attack vector}
We first provide some additional background about $D$'s serialization, i.e., SQLite's database file format~\cite{sqlitedatabasefileformat}.

\begin{figure}
    \centering

\tikzset{every picture/.style={line width=0.75pt}}   

\resizebox{\linewidth}{!}{
\begin{tikzpicture}[x=0.75pt,y=0.75pt,yscale=-1,xscale=1]

\draw    (23,100) -- (106.71,100.71) ;

\draw    (23.71,117.86) -- (106.71,117.71) ;

\draw    (111,136.07) -- (139.03,136.07) -- (139.03,131) -- (157.71,141.14) -- (139.03,151.29) -- (139.03,146.21) -- (111,146.21) -- cycle ;

\draw  (197.2,107.61) .. controls (197.2,103.22) and (200.9,99.67) .. (205.46,99.67) .. controls (210.02,99.67) and (213.72,103.22) .. (213.72,107.61) .. controls (213.72,112) and (210.02,115.55) .. (205.46,115.55) .. controls (200.9,115.55) and (197.2,112) .. (197.2,107.61) -- cycle ;

\draw   (178.64,133.95) .. controls (183.21,133.95) and (186.91,137.51) .. (186.91,141.89) .. controls (186.91,146.28) and (183.21,149.84) .. (178.64,149.84) .. controls (174.08,149.84) and (170.38,146.28) .. (170.38,141.89) .. controls (170.38,137.51) and (174.08,133.95) .. (178.64,133.95) -- cycle (231.69,133.11) .. controls (236.26,133.11) and (239.95,136.67) .. (239.95,141.06) .. controls (239.95,145.45) and (236.26,149) .. (231.69,149) .. controls (227.13,149) and (223.43,145.45) .. (223.43,141.06) .. controls (223.43,136.67) and (227.13,133.11) .. (231.69,133.11) -- cycle (161.98,166.56) .. controls (166.54,166.56) and (170.24,170.12) .. (170.24,174.51) .. controls (170.24,178.89) and (166.54,182.45) .. (161.98,182.45) .. controls (157.41,182.45) and (153.71,178.89) .. (153.71,174.51) .. controls (153.71,170.12) and (157.41,166.56) .. (161.98,166.56) -- cycle (205.46,167.4) .. controls (210.02,167.4) and (213.72,170.95) .. (213.72,175.34) .. controls (213.72,179.73) and (210.02,183.29) .. (205.46,183.29) .. controls (200.9,183.29) and (197.2,179.73) .. (197.2,175.34) .. controls (197.2,170.95) and (200.9,167.4) .. (205.46,167.4) -- cycle ;

\draw  [fill=SpringGreen](248.07,166.56) .. controls (252.63,166.56) and (256.33,170.12) .. (256.33,174.51) .. controls (256.33,178.89) and (252.63,182.45) .. (248.07,182.45) .. controls (243.51,182.45) and (239.81,178.89) .. (239.81,174.51) .. controls (239.81,170.12) and (243.51,166.56) .. (248.07,166.56) -- cycle ;

\draw    (211.69,113.32) -- (224.41,130.4) ;
\draw [shift={(225.61,132)}, rotate = 233.31] [color={rgb, 255:red, 0; green, 0; blue, 0 }  ][line width=0.75]    (10.93,-3.29) .. controls (6.95,-1.4) and (3.31,-0.3) .. (0,0) .. controls (3.31,0.3) and (6.95,1.4) .. (10.93,3.29)   ;

\draw    (198.65,113.32) -- (185.89,131.21) ;
\draw [shift={(184.73,132.84)}, rotate = 305.49] [color={rgb, 255:red, 0; green, 0; blue, 0 }  ][line width=0.75]    (10.93,-3.29) .. controls (6.95,-1.4) and (3.31,-0.3) .. (0,0) .. controls (3.31,0.3) and (6.95,1.4) .. (10.93,3.29)   ;

\draw  [draw opacity=0][pattern=dots, pattern color=Salmon] (451,70.93) -- (451,92.82) -- (450.8,92.82) -- (450.8,143.88) -- (352.94,143.88) -- (352.94,166.84) -- (313,166.84) -- (313,88.71) -- (431.83,88.71) -- (431.83,64.39) -- (450.8,64.39) -- (450.8,70.93) -- (451,70.93) -- cycle ;

\draw  [draw opacity=0, fill=SpringGreen  ,fill opacity=1 ] (451,142.78) -- (451,198.52) -- (357,198.52) -- (357,219.63) -- (313,219.63) -- (313,166.59) -- (431,166.59) -- (431,142.78) -- (451,142.78) -- cycle ;

\draw  [draw opacity=0, fill=SpringGreen  ,fill opacity=1 ] (451,194.02) -- (451,219.5) -- (392,219.5) -- (392,241.71) -- (313,241.71) -- (313,219.5) -- (357,219.5) -- (357,194.02) -- (451,194.02) -- cycle ;

\draw  [draw opacity=0, fill=SpringGreen  ,fill opacity=1 ] (392,219.57) -- (451,219.57) -- (451,241.71) -- (392,241.71) -- cycle ;

\draw    (357,198.52) -- (451,198.52) ;

\draw    (357,198.52) -- (357,219.63) ;

\draw    (357,219.63) -- (313,219.63) ;

\draw    (451,219.57) -- (392,219.57) ;

\draw    (392,241.71) -- (392,219.57) ;

\draw    (258.33,175.48) -- (308.71,49.94) ; 

\draw    (258.33,175.48) -- (308.71,232.87) ; 

\draw    (390,42) -- (390,64.71) ;

\draw    (313,64.71) -- (390,64.71) ;

\draw    (313,88.6) -- (432,88.91) ;

\draw    (432,64.39) -- (451,64.39) ;

\draw    (432,88.6) -- (432,64.39) ;

\draw    (313,166.84) -- (431,166.59) ;

\draw    (353,143.85) -- (451,142.78) ;

\draw   (23,79.86) -- (106.71,79.86) -- (106.71,194) -- (23,194) -- cycle ;

\draw  [draw opacity=0][pattern=north east lines, pattern color=SkyBlue  ,fill opacity=1 ] (450.71,42) -- (450.71,64.71) -- (432,64.71) -- (432,88.6) -- (313,88.6) -- (313,64.71) -- (390,64.71) -- (390,42) -- (450.71,42) -- cycle ;

\draw    (450.71,42) -- (451,241.71) ;

\draw    (313,41.33) -- (313,241.71) ;

\draw    (313,241.71) -- (451,241.71) ;

\draw    (313,41.33) -- (451,41.33) ;

\draw    (390,64.71) -- (313,64.71) ;

\draw    (390,64.71) -- (390,42) ;

\draw    (432,64.39) -- (451,64.39) ;

\draw    (432,64.71) -- (432,88.6) ;

\draw    (313,88.6) -- (432,88.91) ;

\draw (66.19,90.5) node  [font=\footnotesize] [align=left] {Table 1};

\draw (66.64,108.45) node  [font=\footnotesize] [align=left] {Columns};

\draw (66.14,128.45) node  [font=\footnotesize] [align=left] {Row 1};

\draw (65.88,155.45) node  [font=\footnotesize,color=LimeGreen  ,opacity=1 ] [align=left] {Row j};

\draw (65.7,183.45) node  [font=\footnotesize,color=LimeGreen  ,opacity=1 ] [align=left] {Row $\displaystyle j+N-1$};

\draw (65.26,137.5) node   [align=left] {...};

\draw (65.26,164.5) node  [color=LimeGreen  ,opacity=1 ] [align=left] {...};

\draw (204.96,150.25) node [anchor=south] [inner sep=0.75pt]   [align=left] {...};

\draw (243.53,156.48) node  [rotate=-62.25] [align=left] {...};

\draw (167.83,155.98) node  [rotate=-298.68] [align=left] {...};

\draw (184.08,182.86) node [anchor=south] [inner sep=0.75pt]   [align=left] {...};

\draw (227.57,182.86) node [anchor=south] [inner sep=0.75pt]   [align=left] {...};

\draw (206.76,91.45) node  [font=\small] [align=left] {Table 1};

\draw  [draw opacity=0]  (313.03,41.58) -- (390.03,41.58) -- (390.03,64.58) -- (313.03,64.58) -- cycle  ;
\draw (351.53,53.08) node  [font=\small]  {$b_{1} ,\cdots ,\ b_{8}$};

\draw (416,53.16) node  [font=\small]  {$b'_{1} ,\cdots $};

\draw (335.59,75.66) node  [font=\small]  {$\cdots $};

\draw (400.54,76.73) node  [font=\small]  {$\cdots ,\ b'_{2N}$};

\draw  [fill=SpringGreen  ,fill opacity=1 ]  (349.09,142.63) -- (436.09,142.63) -- (436.09,166.63) -- (349.09,166.63) -- cycle  ;
\draw (392.59,154.63) node[fill=SpringGreen  ,fill opacity=1 ]  [font=\small] [align=left] {cell $\displaystyle j+N-1$};

\draw (397.55,208.56) node[fill=SpringGreen  ,fill opacity=1 ]  [font=\small] [align=left] {cell $\displaystyle j+1$};

\draw (421.5,230.14) node[fill=SpringGreen  ,fill opacity=1 ]  [font=\small] [align=left] {cell $\displaystyle j$};

\end{tikzpicture}
}
    \caption{There is one B-tree per database table. Its nodes are stored in individual pages that ultimately form the serialized file. The top-most white area is the header. Next, the blue-lined region contains the cell pointer array. The light pink dotted space is empty, and gets filled bottom-up, as shown by the solid green cells.}
    \vspace{-0.3cm}
    \label{fig:sqlite-serialization-diagram}
\end{figure}
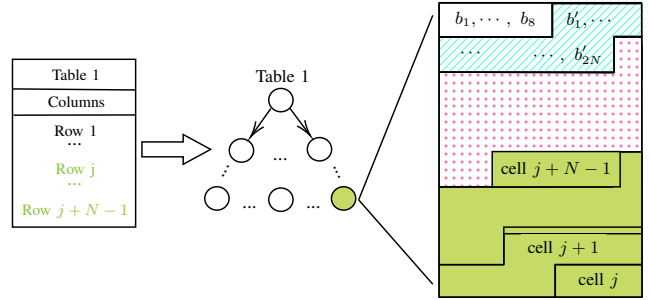

Each table of the database is logically organized as an individual \emph{B-tree}.
The leaf nodes store the actual content of the tables; the internal nodes are irrelevant for our attack,
so we omit their explanation.
Each node is stored in a separate \emph{page}, which are 4,096 bytes in length in the case of WhatsApp (Figure~\ref{fig:sqlite-serialization-diagram}). That is, each page corresponds to only one node of only one table. Then, the serialized file consists mostly of all the pages for all the tables layered sequentially. Indeed, the size of the database file is always a multiple of the page size.

Pages get filled, bottom-up, with \emph{cells}. In particular, leaf pages have one cell for every row in the table, storing the payload and relevant metadata for it. Between headers at the top of the page and the cells at the bottom of it, the rest of the page remains unallocated.

Regarding the content of each cell, consider, for example, the \emph{messages} table of WhatsApp's database, even though what follows applies to any other table. This table has 21 columns, and each of its rows stores relevant information about a single sent or received message. Let $P \coloneqq (p_1, ..., p_{21})$ be one such row.
The structure of its corresponding cell consists of three parts: (1) the size of the encoded row, described below, encoded as a variable-length integer (varint)~\cite{sqlvarint}; (2) the primary key of the row, encoded as a varint; and (3) the actual row $P$, encoded in \emph{record format}~\cite{sqliterecordformat}. The latter consists of at most 21 values denoting the type of each column (e.g., value ``0x07'' means that the type is a 64-bit floating point number), followed by all the $p_i$'s concatenated together, encoded in some predefined manner (e.g., text in UTF-8). In particular, one of these $p_i$ contains the actual text content of the message.

With this additional context, we can see that, if two messages are sent close to each other in time, the cells for their resulting rows in the \emph{messages} table (and all other tables that include message contents) will lie close to each other in the serialization $S$ of the database $D$.  In particular, unless the messages are very long, they will lie within the same zlib search window of 32,768 bytes. Then, when the next backup is triggered and zlib is run, repeated substrings in $S$ get replaced with short references to their prior occurrences, if any.  Thus, if some message $m$ contains (parts of) some other message $s$, we expect the output of zlib to be shorter than if~$m$ does not contain $s$: in the former, both cells will contain repeated text content, which zlib compresses. This output is finally encrypted with AES-GCM, whose output reveals the length of the (compressed) input plaintext. We note that zlib does not compress repeated strings unless their length is greater than or equal to four bytes, so this is a lower bound on the number of consecutive bytes that $m$ and $s$ must match on. 

\paragraph{General attack structure}
This logic can be exploited as a message recovery attack, in the quiet device threat model: for some $\calD \subseteq V$, where all strings in $\calD$ are of the same length, the adversary seeks to determine which $v \in \calD$ is present in the database. To do so, after the victim receives the target string from some third party, the adversary injects every string in $\calD$ as a standalone message, interleaved between backups, and measures the increase in database size to determine which message contains the target string. This method requires $|\calD| + 1$ backups in total (the additional backup is required to measure the initial size of the database).

Instead of this brute-force injection strategy, an adversary can use a binary injection strategy to decrease the number of required backups. In this case,  the adversary first splits $\calD$ into two subsets of equal size, $\calD_0$ and $\calD_1$, and sends \emph{all} of $\calD_0$ and $\calD_1$ as two separate messages interleaved between backups. Conceptually, this is a distinguishing attack, between $\calD_0$ and $\calD_1$. Then, whichever half contains the target string will likely result in more compression than the other, i.e., a smaller difference in the size of the encrypted backups that were triggered before and after sending the message containing the matching string. So, they can then make a recursive call to this algorithm, using this half as the input set. The adversary continues this procedure, until the input set has just one string in it, which is deemed the found string. This method requires $2 \cdot \lceil \log_2 |\calD| \rceil + 1$ backups in total, since each iteration reduces the search space in half.

One issue with binary injection is that each candidate string is inserted multiple times instead of once. However, we can improve this strategy by sending different (potentially overlapping) \emph{substrings} of the candidate string in the various recursive calls. For example, if one potential target string is  ``abcdefgh'', in the first and second recursive calls we can instead send ``abcd'' and ``efgh'', respectively. That way, both of them will \emph{only} match the target string.

In this strategy, the length of the substrings and how much they overlap are tunable parameters, which depend on the lengths of the candidate set~$\calD$ and the target string~$v$. The adversary can determine empirically, with local testing, which ones achieve the highest probability of success for the set of candidates (which they know a priori). These substrings, however, must be at least 4 bytes long, as discussed above. We note that this optimization is not applicable if the number of candidate strings is significantly larger than the length of the target string.
In particular, the minimum requirement---which occurs when the length of the substrings is 4 bytes and their overlap is 3 bytes---is $\log_2|\calD| \leq |v| - 3$. So, the adversary can use the optimized variant of the binary search attack if these preconditions are satisfied, or else use the normal version instead.

Different injection strategies bound the size of $|\calD|$ and the strings therein, as they must all lie inside zlib's search window of 32,768 bytes. As a concrete example, the minimum requirement for the brute-force strategy is that $(|v| + \epsilon) \cdot |\calD| < 32,768$, where $\epsilon$ is the additional number of bytes in the smallest cell containing $v$, across all tables. By measuring the sizes of cells in decrypted databases for known payloads, we estimate that $\epsilon$ is roughly $13 (\pm 5)$ bytes. Conversely, for the binary injection strategy, the various recursive calls yield a geometric sequence with common ratio of $1/2$, which results in a smaller upper bound on $|\calD|$ than for the brute-force variant. Solving this sequence yields that the minimum requirement is now $n \cdot (\frac{2^n-1}{2^{n-1}}) < 32,768 \implies n < 16,384$, where $n = (|v| + \epsilon) \cdot |\calD|$.

\paragraph{Sources of noise}
The attack may fail due to the fact that sending a message inserts more than just its text content into the database, such as timestaps and other metadata. All of this also gets compressed, which adds noise to the measurements. Empirically, this type of noise has relatively consistent impact on compression across messages, so it results in a similar increase in size across all measurements. Note that the impact of this noise in the probability of success is a function of the length of the strings in $\calD$, as compression from shorter strings will be too obfuscated by this noise.

An interesting idea for future work would be to use statistical techniques to mitigate some of these limitations, such as in~\cite{al2013lucky,zindros2016practical,hogan2023dbreach}. For example, as $|\calD|$ or $|v|$ increase, the adversary might a priori record the intrinsic compressibility of each message, and ``penalize'' those that have higher compression, in order for this not to bias the measurements.

Lastly, we restate that this proof-of-concept attack only works in the quiet device setting; other messages received by the victim might push the target string outside of the compression window or match injected content, making the attack fail.

\paragraph{Experimental evaluation}
\begin{figure*}
    \centering
    \renewcommand{\arraystretch}{1.1}
    \begin{tabularx}{\textwidth} { 
         >{\raggedleft\arraybackslash}X| 
         >{\raggedleft\arraybackslash}X 
         >{\raggedleft\arraybackslash}X 
         >{\raggedleft\arraybackslash}X
         >{\raggedleft\arraybackslash}X|
         >{\raggedleft\arraybackslash}X 
         >{\raggedleft\arraybackslash}X 
         >{\raggedleft\arraybackslash}X
         >{\raggedleft\arraybackslash}X}
     \toprule
     & \textbf{$|v| = $ 4} & \textbf{12} & \textbf{20} & \textbf{28} & 
    \textbf{4} & \textbf{12} & \textbf{20} & \textbf{28}\\
      \midrule
      $|\calD| = 2$ & 74 & 83 & 97 & 100 & 74 & 83 & 97 & 100 \\
      4 & 44 & 73 & 99 & 99 & 64 & 92 & 99 & 99\\ 
      8 & 38 & 62 & 82 & 92 & 22 & 33 & 42 & 50\\
      16 & 14 & 50 & 76 & 81 & 7 & 10 & 16 & 22\\
     \bottomrule
\end{tabularx}

\caption{Experimental success probability (standard deviation of $\pm 2\%$) for the message recovery attack (\secref{sec:compression-attacks}) in our local simulation. Rows correspond to different values of $|\calD|$, columns to different values of $|v| \in \calD$, and cells the probability of success. The left sub-table corresponds to the brute-force injection strategy, and the right one to the binary injection strategy (note that these are equivalent when $|\calD| = 2$).}
\vspace{-0.3cm}
\label{fig:compression-attacks-experiments}
\end{figure*}

We experimentally verified both the brute-force and binary injection attacks. To do so, we set up two testing environments: real WhatsApp clients on Android phones, and a local simulation of the core client-side operations relevant to the attacks.

The local simulation operated on WhatsApp databases that were downloaded beforehand and decrypted locally, to obtain their corresponding database files. Then, the environment consisted of two parts: (1) a limited API to make all changes to the database file that occur in the real client after a new message is received, and (2) a compression and encryption pipeline to generate the backup ciphertexts. We describe the details of how we reversed engineered these steps in Appendix~\ref{sec:a-experiments-local}, and note that this simulation-based approach is not uncommon when evaluating deployed systems, e.g.~\cite{backendal2023mega, patersonthree}.

We used the real environment to implement the attacks and run a limited number of trials to confirm their overall correctness. Then, we used the simulated environment to run various independent trials of these same attacks, and empirically estimate the probability of success. This local setup allowed us to (i) run thousands of experiments in a reasonable amount of time and without overloading WhatsApp's servers; and (ii) run various independent trials of the same experiment, starting with the same state.

We measured the performance of each attack on different values of $|v_b|$ and $|\calD|$, in the quiet setting, to gauge how the probability of success degrades as strings get shorter and/or there are more candidate strings. Each experiment required $|\calD| + 1$ and $2 \cdot \lceil \log_2 |\calD| \rceil + 1$ backups for the brute-force and binary injection strategies, respectively.

For all strings in $\calD$ (across all tests), we used alphanumeric strings sampled uniformly at random. For each $(|v_b|, |\calD|)$ pair, we ran 1,000 independent trials, recorded the number of these in which the attack successfully recovered the string, and deemed this the success probability. The results are displayed in Figure~\ref{fig:compression-attacks-experiments}. We emphasize the first row of the table, when $|\calD| = 2$, which represents the particular case of a distinguishing attack (in which case both injection strategies are equivalent).

In practice, the composition of the strings plays a role in the probability of success of the attacks: if the strings are very similar, their impact on the size of the ciphertext is also very similar.
Therefore, we additionally ran our attack on a more restricted corpus of strings, to simulate how it can be used to retrieve sensitive data of real-world interest. To do so, we used sample Social Security Numbers (SSNs), credit card numbers (CCNs), and passwords. SSNs were sampled uniformly at random from the set of possible SSNs, in accordance with the issuance guidelines from the Social Security Administration~\cite{ssnguidelines}. CCNs were sampled at random in accordance with basic Visa and Mastercard guidelines and the Luhn formula~\cite{enwiki:1160354109}. Lastly, passwords were sampled uniformly at random from a popular list of 10,000 common user passwords~\cite{githubcommonpasswords}. We show the results of these experiments in Figure~\ref{fig:compression-attacks-experiments-real-data}.

\begin{figure}[t]
    \centering
    {\color{spdiff}
    \begin{tabularx}{0.47\textwidth} { 
         >{\raggedleft\arraybackslash}X| 
         >{\raggedleft\arraybackslash}X 
         >{\raggedleft\arraybackslash}X 
         >{\raggedleft\arraybackslash}X}
     \toprule
     & SSNs & CCNs & Passwords \\
      \midrule
      $|\calD| = 2$ & 81 & 97 & 89\\
      4 & 47 & 49 & 45 \\ 
      8 & 40 & 42 & 38 \\
      16 & 31 & 25 & 20 \\
     \bottomrule
\end{tabularx}

\caption{Experimental success probability (standard deviation of $\pm 2\%$) of our message-recovery attack for different target types---Social Security numbers (SSNs), credit card numbers (CCNs), and common passwords---using our brute-force injection strategy.}
\vspace{-0.3cm}
\label{fig:compression-attacks-experiments-real-data}}
\end{figure}

Our attacks are successful by standard cryptographic measures: they performed noticeably better than a random guess, i.e., success probability greater than 50\% for distinguishing attacks, and $|\calD|^{-1}$ for message recovery attacks. However, we can see that the probability of success degrades as $|\calD|$ increases, which also requires more backups.

\subsection{Exploiting the Keyword Search Index}~\label{sec:ftsattack}
The prior two attacks exploit deduplication and zlib compression before encryption. A natural countermeasure would be to simply turn off both forms of compression. However, this is not sufficient to prevent injection attacks, due to subtleties in how SQL processes and stores message
data sent from different users.

In this section, we consider a modified version of WhatsApp that turns off zlib
compression, and describe a
distinguishing attack that exploits cross-user interactions in a keyword search
index contained within the SQL database~$D$. The attack determines which of two
messages was sent by another honest user $U'$ to the target user~$U$.  The
attacker needs to observe at most 16 backups, and inject at most 44 messages. The reason for the modified setup is that this attack is more complicated than the distinguishing
attack of Section~\ref{sec:compression-attacks}, so in this section we will
assume, for the sake of argument, that compression is no longer being used (otherwise, an attacker would just opt for the simpler, zlib-based
distinguishing attack to achieve the same goals).

The attack proceeds in several phases, and assumes that (1) the attacker can
sandwich $U'$'s message with adversarial messages, and (2) we are in the quiet device setting, i.e.,~$U$ receives no other
messages beyond the target and adversarial ones while the attack is running. As
such, it is arguably less practical than the attack exploiting compression:
WhatsApp's daily backups would imply that our attack requires the target client
to be ``quiet'' for up to 16 days. Nevertheless, it is a proof-of-concept that
compression is not the only vector for information leakage in injection attacks. This attack is a symptom of a higher-level issue: injection attacks may
arise from subtle cross-user interactions in internal data structures of the
database.

\paragraph{The FTS4 keyword search index} WhatsApp uses SQLite's
FTS4~\cite{sqlitefts4} module to implement full-text search on all sent
and received messages. The attack depends on many low-level details of FTS4, so we
give some background.

An FTS4 index is logically structured as an inverted index, mapping terms to the
documents (messages) in which they appear. After a message $m$ is received, it is
processed to extract the terms $t \coloneqq (t_1,
..., t_r)$ it contains. This process---tokenization---is application-specific.
For WhatsApp, $m$ is first cast into all lowercase, and all non-alphanumeric
characters are replaced with whitespace. The resulting words are the terms. For
instance, the message ``They've paid \$1!'' would add ``they'', ``ve'', ``paid'', and ``1'' to the
index. Each message has an associated document ID (\docid), 
assigned sequentially as messages arrive.

The entire index is organized as a collection of independent B-trees: their leaves consist of a sequence of
term, document list (from now on, doclist) pairs. Each entry in the latter specifies the \docid (and the
position of the text within the message) in which the term appears. The pairs in a
single leaf are arranged based on the lexicographic ordering of the terms. We have omitted some
lower-level details about term-doclist pairs for simplicity, and
refer readers to~\cite{sqlitefts4} for these. Given a sequence of tokenized
messages~$R$ (with their associated \docid's), encoding them in a leaf node as
term-doclist pairs is deterministic; we denote the process by~$\node(R)$.

The B-trees are organized into levels. Every time a new message is inserted into
the index, a new level-0 B-tree is created, containing only the terms in this
message, each mapping to the same \docid. After 16 level-0 B-trees are inserted, they are
merged together into a new level-1 B-tree, which follows the same structure as
before, except that it now contains the terms of 16 messages. The value of 16 here is
configurable and represents a trade-off between insertion speed and search
speed; WhatsApp uses 16.  In general, after enough level $i$ B-trees have
accumulated, a new level $i+1$ tree gets created, merging the contents of all
prior level $i$ trees and deleting them. Thus, as the index grows, it will
consist of multiple B-trees of different sizes, all of which must be scanned
when performing a search. Note that this implies that a term, albeit unique
inside \emph{each} tree, may be repeated across \emph{different} trees, which
get consolidated when/if these trees merge.

We now discuss how these multiple B-trees get incorporated into the
serialization $S$ of the database $D$.
An FTS4 index is supported by five tables.
The table most relevant to our
attack is \emph{messages\_ftsv2}$_{segdir}$, which serves as a ``directory''
for all B-trees of the index. This table has one row per tree, which contains
its level, index within the level, some metadata about its nodes (e.g., ID of
the first and last leaf nodes, its level, etc.), and, most importantly, the entire
root of the B-tree. All other non-root nodes, for all trees, are stored in a
separate table. Note that, if the tree is not too big, it may be the case that
it fits entirely within the root node, in which case the entire tree is stored
in \emph{messages\_ftsv2}$_{segdir}$. Indeed, our attack only requires small
messages, satisfying this precondition. So, the takeaway from this is that, for our
purposes, all trees of the index are stored in \emph{messages\_ftsv2}$_{segdir}$
as a single root node (with additional metadata).

Recall from our prior discussion on SQLite serialization
(\secref{sec:compression-attacks}) that tables are also logically
organized as B-trees, and each page in the serialized file corresponds to one of
the nodes. This means that \emph{messages\_ftsv2}$_{segdir}$ is itself a single
B-tree, with its various nodes encoded as pages in~$S$. So, the
B-trees from the FTS index will be stored as cells in the leaf pages of
\emph{messages\_ftsv2}$_{segdir}$. This means that \emph{messages\_ftsv2}$_{segdir}$ is a
B-tree containing yet more B-trees (from a different context) in its leaves.

To summarize, the main steps (for our purposes) when~$U$ receives a message $m$ are as follows:
\begin{newenum}
    \item $m$ is tokenized to extract the terms it contains.
    \item A level-0 B-tree gets created for this message, mapping all its terms
      to its \docid. Assuming the message is not large, this tree consists of a
      single (root) node. This could trigger one or more merges to higher levels.
    \item A row containing this root node gets added to \emph{messages\_ftsv2}$_{segdir}$.
    \item Under-the-hood, the prior step adds a new cell to the last leaf page of \emph{messages\_ftsv2}$_{segdir}$'s B-tree.
\end{newenum}
For a sequence of term-doclist pairs $R$, we use $\cell(\node(R))$ to denote the
cell in the leaf page of \emph{messages\_ftsv2}$_{segdir}$ that stores the tree
containing $R$ (in our attack, the tree is always a single node). Other
contents of the row (e.g., the level of the tree) are left implicit.

Importantly, knowledge of $R$ and its ``maturity'' (its level and age within the
level) is sufficient to deterministically compute $\node(R)$. Further,
knowledge of $\node(R)$ and its primary key in the table are enough to construct
$\cell(\node(R))$. Since the encoding of the primary key is at most eight bytes (as a varint), the cell's
size can be estimated very closely even without knowing the primary key, i.e.,
just with information about~$R$. 

\paragraph{Attack idea}
Sending a message containing a string~$s$ can modify the index in one of two ways: (1) if $s$ was
not present in some other message, a new entry in the index gets created,
mapping $s$ to (only) the \docid of this message; (2) if $s$ was present in some
other message, the ID of this message gets appended to the (already existing)
entry for $s$. Since the first scenario results in a more substantial
modification of the database, this is a potential source of leakage.

Exploiting this idea is challenging, 
for two main reasons. First, since every new message creates a new
level-0 B-tree, the cross-user interaction will not be detectable until the message
containing~$s$ and the target message are merged together within a higher-level
B-tree.
So, the adversary needs to ``trigger'' this merge.

The second main challenge is that the size of the serialized database file is
always a multiple of the page size (4,096 bytes in the case of WhatsApp). So,
unless the strings are sufficiently large, there is not enough granularity in
the size of the database to meaningfully distinguish small differences in the
size of the index. In the prior attacks, zlib would prune out the empty space,
but here we assume zlib is turned off.

To deal with the first challenge, 
our attack ``isolates'' a level-1 B-tree containing
the target string and one of the two candidate strings in a fresh leaf page of
the \emph{messages\_ftsv2}$_{segdir}$ table that only contains
adversarially-chosen payloads. Then, the adversary can use injections to 
measure the amount of
empty space to determine the size of the tree (and, thus, if the two strings got
stored together~or~not).

\paragraph{Attack for a simplified setup} Let's start with a simplistic example
to show the core ideas of the attack---moving forward, we refer to this variant
as the simple setup. Assume that the target string is ``my password is foo'',
and that it is the first and only message in $D$ (so, its \docid is 1). This means
that, to start, the index only has one level-0 B-tree, containing term-doclist
pairs [(foo, [1]), (is, [1]), (my, [1]), (password, [1])], all in a single root
node (other low-level details omitted for clarity). 

The adversary sends a message containing ``my password is~$m$'', for some guess
$m$, which creates an analogous level-0 B-tree. To trigger the cross-user
interaction, the adversary then sends 15 arbitrary messages, say the character
``a'' for all. Each of these also creates an individual level-0 B-tree, e.g.,
[(a, [7])]. The last of these messages triggers a \emph{merge}, given
that there were already 16 level-0 B-trees present in the index. So, all these
trees get deleted, resulting in a new level-1 B-tree that precedes the last
level-0 B-tree (the one that triggered the merge). Internally, both
trees will be stored as cells in the bottom of the first leaf page of the
\emph{messages\_ftsv2}$_{segdir}$ table (with the level-1 tree's cell below the
level-0's one). The rest of the page, as described in
Section~\ref{sec:compression-attacks}, is unallocated for now.

Assume that $m$ = ``foo''. Then, the new level-1 B-tree will contain [(a, [3,
..., 16]), (is, [1, 2]), (foo, [1, 2]), (my, [1, 2]), (password, [1, 2])].
Conversely, if $m$ = ``bar'', there are (only) two changes to the tree: it has a
new term-doclist pair, (bar, [2]), and ``foo''s entry is just (foo, [1]).
The first case---a correct ``guess''--- results in a smaller level-1
B-tree ($T'$) than the second case ($T''$), since the former only has one of the
two candidate strings in it. So, there are two options for the amount of empty
space in the leaf page of the table: either $N' \coloneqq 4096 - |T'| - |B| - 12$ or
$N'' \coloneqq 4096 - |T''| - |B| - 12$. Here $B$ is the (single) level-0
B-tree that triggered the merge and 12 is the length of a header.

Then, the last step is to measure the amount of empty space in the page to tell
which of the two scenarios the data structure is in. To do so, the adversary can
progressively fill up the page with new (cells of) messages, until a new 4096-byte
page gets triggered. Since the adversary knows the size of the messages, they can
work backwards to estimate the amount of empty space. In particular, the last
payload that triggers the new page gets added to that new page.

Note that the adversary can only measure this once: after a second page is
allocated (e.g., if they overshot the first measurement), they cannot revert to
the prior page and measure again. 
This rules out most simple measurement techniques, such as 
binary search on the amount of empty space. Instead, the adversary can leverage
the fact that they know the lengths of all strings in the page. So, they can
locally compute $N'$ and $N''$, and send a single message whose payload is between
these two options. This confirms if the strings matched (no new page is
allocated) or not (new page is allocated). This simple example requires a single
backup, to know by how much $|S|$ grew.

The goal of the phases that follow is to adapt this simple setup to the setting where the initial state of the B-trees could be anything, i.e., where the victim has received some unknown number and set of messages from honest parties before the attack begins.

\paragraph{Full attack} Let us now move to the more general case, where the
adversary knows nothing about the prior messages~$U$ has received. This prevents
the simple setup attack above from working: as we just saw, the adversary's measurements depend on knowing the
exact contents of the B-tree where the target string is present. Thus, the goal of our full attack is to adapt the simple setup to the setting where the initial state of the B-trees could be anything, i.e., where the victim has received some unknown number and set of messages from honest parties before the attack begins. Our attack proceeds in several phases:
\begin{newenum}
\item \emph{Flush level-0 B-trees:} The adversary sends a sequence of 16 messages $c_1,\ldots,c_{16}$ to
  ensure that all level-0 B-trees contain only
 adversarial content.
\item \emph{Push to a new page:} The adversary waits for a backup, measures
  its length, sends a large message~$M$ (e.g., $|M| = 4000$ bytes), and
  waits for another backup. The next backup should be larger, meaning that $M$
  forced allocation of a new page. If the size of the database did not change, the adversary can send a second large string without jeopardizing the rest of the attack.
\item \emph{Wait for the target message:} The adversary waits for an honest
  party to send a message $v_b \in \{v_0,v_1\}$ to $U$. This adds a level-0 B-tree containing just $v_b$.
\item \emph{Send a guess:} The adversary sends $v_0$ as a guess to $U$. This
  adds another level-0 B-tree, which contains just~$v_0$.
\item \emph{Trigger a merge:} The adversary sends 13 messages 
  $c_1',\ldots,c_{13}'$ to ensure that a merge
  occurs, which combines the level-0 B-trees containing $M$, $v_b$, $v_0$ into a
  single level-1 B-tree. This B-tree contains some unknown number of the 
  flushing messages sent in step (1), depending on the number~$n$ of level-0 B-trees present in the
  system before the attack started. The adversary does not know $n$.
\item \emph{Measure:} The adversary iteratively sends a sequence of
  ``measurement'' messages~$m_1,\ldots,m_{13}$, waiting for a backup in between each sent message. 
  When the adversary detects that a new page is allocated, it can infer the bit~$b$
  as the number of messages sent mod 2. This last measurement is quite delicate,
  as it relies on arranging that the lengths of the $c_1',\ldots,c_{13}'$ and
  the lengths of the $m_1,\ldots,m_{13}$ are such that as long as $0 \le n < 14$
  the number of measurement messages reveals $b$.
\end{newenum}

The above glosses over a number of subtleties that we unpack in the extended version of the paper. 

In total, the attack requires at most 44 adversarial messages 
and 16 backups. All messages, besides $P$, are small (and, in fact, need to be for the
attack to work). For all of them, except for $v_0$, only the length matters, and the content is irrelevant. The attack requires that the byte length of the candidate strings
$v_0,v_1$ are approximately between 40 and 2,000 bytes (approximately because of
some variation in header lengths; see the extended version of the paper).
Furthermore, the attack can fail for two reasons. First, it fails should $n \in
\{14,15,16\}$. 
Recall that~$n$ is based on the 
number of messages sent and received by $U$ before the attack; under the assumption that
this is uniformly distributed, this failure only arises about 3/16 of the time. 
One could correct for this failure at the cost of a more
complex and expensive preparatory phase, but we did not implement this. 

Second, depending on the state of the target client, it might be that the attack
triggers a merge of level-1 trees into a level-2 tree, thereby obviating the
attack's goal of isolating the target message in a level-1 tree with just
adversarial data.  This will not happen if the adversary's
messages are the first ever to be sent to the target. In the steady state, and
assuming the number of received messages is uniformly distributed, the
probability of this failure is roughly 1/256 (since a level-1 merge occurs every
256 messages).
We refer to other technicalities related to this attack in the extended version of the paper.

\paragraph{Implementation} Unfortunately, since this attack assumes a modified
version of WhatsApp, we were not able to deploy it on real WhatsApp clients.
Instead, we used our local simulation once again
(\secref{sec:compression-attacks}), where we could turn off
compression to achieve the desired setup. All other aspects of the simulation
testing environment were left unchanged.

In this setup, we successfully implemented the attack for varying $n \in
[1\dotdot 13]$ and testing both the $b=0$ and $b=1$ cases, for randomly chosen
$v_0,v_1$. Note that the composition of the strings and their
length do not affect the performance of the attack, as long as they are within
the specified bounds (unlike for the compression attack).  As before, we
used real WhatsApp databases, downloaded and decrypted locally, as inputs to the
attack.

\section{Injection Attacks against Signal}
\label{sec:passive-attack}

We now turn to Signal, which has a significantly different backup approach.
While we have not found attacks that recover message contents, we
demonstrate injection attacks that reveal potentially sensitive metadata such as the number of contacts and
messages received by a target user~$U$.
We will start by describing Signal's backups more in detail, then present an attack in the quiet device setting, and finally extend it to the noisy device setting.

\paragraph{Signal's backup structure}
When building a backup, Signal creates a representation of the tables in its
SQLite database~$D$ by generating SQL statements suitable for recreating the
database: a {\sf CREATE} SQL statement for all the tables and an {\sf INSERT} SQL
statement for each row in the table. SQL statements are encoded as byte strings
using the protobufs library~\cite{protobufs}.
For media objects, such as videos or stickers,
Signal creates a media metadata header, which contains the length of the
media object and other information, and serializes it via the protobuf library~\cite{protobufs}. This results
in a sequence of plaintext byte strings, called frames: 
\begin{newmath} 
  pb_1\:,\:pb_2\:,\:\ldots\:,\:mh_k\:,\:f_{k+1},\ldots
\end{newmath}%
where $pb_i$ represent protocol buffer values for SQL statements, $mh_i$
represent media metadata header protocol buffer values, and $f_i$ represent
media file contents. 

This structured plaintext is then component-wise encrypted as follows. 
The 30-digit uniformly chosen passphrase~$P$ is run through a key derivation
function (HKDF~\cite{hkdf}) with a fresh salt $\salt$ to generate a secret encryption key (for AES-CTR), and a secret
authentication key (for HMAC).  It also generates a random 
initialization vector $\IV$ for AES
use with AES-CTR mode.
Then all the frames are encrypted and
authenticated separately (using the same $\IV$ but incremented
appropriately). For SQL statement and media header frames, a four-byte
length is prepended to the corresponding ciphertext. This step is skipped for
media file contents, since the length is in the preceding (encrypted) frame.
Thus, the resulting backup ciphertext has the following form:
\begin{newmath} 
  C_0 = 
  \ell_1 \concat \calE(pb_1) \concat \ell_2 \concat \calE(pb_2)\concat \cdots \concat \ell_k \concat \calE(mh_{k})\concat \calE(f_{k+1}),\ldots
\end{newmath}%
where the $\ell_i$ values are four-byte plaintext encodings of the length field for the subsequent
ciphertext and $\calE$ represents the authenticated encryption processing (where
we omit keys and IV from the notation for simplicity). Prepended to this is a
header including a plaintext length field followed by~$\IV$ and~$\salt$.

We note that frames are deterministically ordered given an input~$D$: first are all the {\sf CREATE} SQL statement frames for all the tables, next the {\sf INSERT} SQL statement frames, first for the \emph{contacts}, \emph{threads} (a table containing the last message of every chat), and
\emph{messages} tables, and then the rest of the tables are ordered based on their names alphabetically.
We will focus on the first three tables, which we denote by $T_{contacts},
T_{threads}$, and $T_{messages}$, respectively.
Insert statements for each table are handled in increasing order of their row number. Additionally, for a row that has an associated media file, the ciphertext from the media header and the media file frames will be concatenated after the row.

This structured ciphertext is almost completely parsable into separate component
ciphertexts by someone without~$P$.
The only challenge is that media frame lengths are hidden, and it is unclear where these frames end. As specified before, $T_{contacts}, T_{threads}$ and $T_{messages}$ appear before any media frames. This makes parsing the
frames for these tables straightforward, as one just follows the length fields
starting from the first header length field.  
Parsing without $P$ beyond tables with media frames can be accomplished heuristically
looking for four-byte sequences that are likely to be length fields (they have many
high order zero bits). We will not need this extension for the attacks below,
which focus on the first three tables.

\subsection{Injections in the Quiet Device Setting}
Given $C_0$
the adversary learns a sequence of lengths $L_0= \ell^0_1,
\ell^0_2, \ldots$ but is unsure which frame lengths correspond to which tables,
since, a priori, they do not know how many rows are in each table.
To disambiguate we can use an injection attack. 

For the attack we will start in the quiet device setting and
no noise, and we will extend it to the noisy device setting in the next subsection.
Under these assumptions,
after observing $C_0$, an adversary can use a phone number that has
never interacted with~$U$'s account previously and sends a message~$m$ from
the new number to~$U$. Then the adversary waits for a backup to obtain a new
ciphertext $C_1$. Assuming there is no noise, i.e., none of the rows in $C_0$ changes between $C_0$ and $C_1$, the adversary can parse out a sequence of lengths $L_1 = \ell^1_1, \ell^1_2, \ldots$ and look for
the first location $\ell^1_i$ where the
sequences differ ($\ell^1_i \not= \ell^0_i$).
This will be the new frame due to the injected content within
the first table $T_{contacts}$. The second difference
($\ell^1_{j} \not= \ell^0_{j-1}$) will be for
$T_{threads}$ and the third ($\ell^1_{k} \not= \ell^0_{k-2}$)
for the $T_{messages}$.
A problem occurs should the length of one of the three new frames
resulting from the injected message 
have a length that collides with one of the frame lengths to the left or right in the sequence. 
The adversary can avoid this by choosing the length of~$m$
appropriately. 

Ultimately this reveals the exact number of rows, and each of their
corresponding SQL statement sizes, for at least the first three tables. This at least allows counting the number of distinct contacts and received messages from $U$.

\paragraph{Experiments}
To check that this approach works as expected, 
we tested the attack using three Android Pixel 4
emulators, as described earlier, to simulate the victim ($U$), the
adversary ($\mathcal{A}$), and an honest party ($U'$) that might communicate
with $U$.
We used Signal v6.22.8 and performed
the following experiment eight times.

We first created new Signal accounts for the three devices, and
ensured $\mathcal{A}$ was not included among $U$'s contacts. Then, we selected
ten operations to be performed between $U$ and $U'$, to create a
randomized starting state. We randomly sampled among the following actions and
weights: send message (15$\%$), receive message (15$\%$), send image (15$\%$),
receive image (15$\%$), call (10$\%$), receive call (10$\%$), create group
(10$\%$), delete message (5$\%$), and delete group chat (5$\%$). In case a delete
operation was picked and there was nothing to be deleted, we resampled the
action. Group names and message's length and content were selected at random.

We then executed the actions in the order sampled, having
only one device active at a time. Afterwards, we enabled backups in $U$
and created a backup $C_0$. Then, for the injection, we selected a random
string and sent it from~$\mathcal{A}$ to $U$. Finally, we
created a second backup $C_1$ from $U$. Importantly, to recreate a
quiet device setting, $U'$ was turned off between the two backups.

We scanned both backups for their frame boundaries and
created two lists $L_0$ and $L_1$ containing the sizes of the frames in $C_0$
and $C_1$, respectively. We then scanned the two lists to find the first
three differences, as described above, and output the estimated length of the
three tables. To verify success, we decrypted $C_1$ and confirmed the
sizes of the tables. The attack always succeeded.

\subsection{Extension to the Noisy Device Setting}
In the noisy device setting, the previous strategy would
fail since any change in $U$'s database, whether it is receiving
a new message or simply changing the name of a contact, would
create or modify a frame. 
To overcome this challenge, we can leverage the fact that the 
adversary controls the size and/or number of the injections.

Whenever a user receives a message, the respective rows in $T_{message}$
and $T_{threads}$ will store at most 2,000 characters of the message received.
The rest will be moved to the table containing media attachments. But this still
means that large injected messages will stand out from shorter messages, e.g.,
ones that are at most 1,900 bytes. For short noise workloads, $\advA$ can
simply inject very long messages before a backup, observe the backup to retrieve
a list of frame lengths $L_0 = \ell_1,\ell_2,\ldots$, and then look for the first two values $\ell_i > 2000$ and $\ell_j > 2000$. 
This relies on frames in $T_{contacts}$ having length less than or equal to the
threshold 2,000; this held true in all our experiments. 
Then $T_{threads}$ ends at
location $i$ and $T_{messages}$ ends at location~$j$. Note that this requires only one
backup. 

We experimentally checked that this works in the same setup as the quiet device
setting above, except that we injected a message of maximum size. We then
checked in the resulting backup $C_1$ that the first two frames of length larger 
than our threshold corresponded to the injected message's addition to
$T_{threads}$ and $T_{messages}$.  

The above strategy can be refined to handle noise from long messages by having 
the adversary inject a sequence of messages with different lengths that are,
with high probability, guaranteed to be distinct from the noise. More
specifically, we observe that injecting a message of length~$\lambda$ bytes leads
to a frame length of $\ell = \lambda + \delta_1$ in $T_{threads}$ and $\ell' =
\lambda + \delta_2$ in $T_{messages}$ where $\delta_1,\delta_2$ both vary in our
experiments across insertions---we are not sure what dictates their exact values. 
But in our experiments so far, both $\delta_1$ and~$\delta_2$ fall in the range
$[447\ldots 465]$. 

Despite this uncertainty, the adversary can use the header lengths as a noisy channel through
which the adversary sends a sufficiently long uniformly random message. Towards
this, let~$\deltamin,\deltamax$ be the minimum and maximum possible~$\delta_1$ and $\delta_2$. 
Also, let $r~=~2000~/~(\deltamax~-~\deltamin~+~1)$.
Then the adversary~$\advA$ injects 
a sequence of~$m$ messages of lengths defined as follows. First choose
$x_1,\ldots,x_m \getsr [1,r]$. Then, inject for the $i\thh$ message 
one whose length is $\lambda_i = x_i(\deltamax-\deltamin+1)$. 
Upon observing the backup, $\advA$ looks for a sequence of $m$ lengths
such that the $i\thh$ length $\ell_i \in [\lambda_i +
\deltamin,\lambda_i+\deltamax]$.  This will reveal the
end of $T_{threads}$, and the adversary can look for the pattern again to find
$T_{messages}$. 
Assuming that there are at most~$\nu$ frames added to the backup before the
injection (e.g., due to other messages received by the target), then via a union
bound, the probability the attack fails is at most
$\nu\left(\frac{1}{r}\right)^m$. For $[\deltamin,\deltamax] =
[447,465]$, setting $m = 4$ ensures that the probability of
failure is at most $2^{-13}$ even for $\nu$ up~to~$10,000$. 

Finally, we note that these approaches do not reveal the size of $T_{contacts}$
or $T_{threads}$ because they do not detect the end of $T_{contacts}$. We
believe one could extend our techniques to do so with high probability of
success, but have not explored this further.
\section{Discussion}\label{sec:discussion}

\paragraph{Practicality of the attacks} We first discuss the practical viability of
our attacks, which varies across the four we introduced (see
again \figref{fig:attacks-summary} in the introduction).  As discussed in
\secref{sec:architecture}, our attack setting requires access to
the backups, which in turn means that the adversary has already been able to
compromise a user's cloud backup account for
WhatsApp or, for Signal, wherever backups end up stored. This may be difficult
for many adversaries, and ensuring security of the backup storage prevents
our attacks.

Another practical issue is noise, and our attacks have varying levels of
robustness to it.  The WhatsApp attachment
dictionary attack and Signal metadata inference attacks work despite many kinds
of noise, and therefore we expect they will work even if typical, active use of
the messaging application by the target occurs during the attack.
In contrast, the WhatsApp attacks
exploiting zlib and FTS4 are fragile to noise and
therefore we expect that they may be hard to mount in practice. 
Even so, future attacks may offer improved robustness, and 
achieving stronger end-to-end security for backups suggests we need to seek mitigations that remove, or hinder exploitation of, the vulnerabilities 
underlying our attacks.

\paragraph{Comparison between WhatsApp and Signal} 
Towards mitigations, we first observe that the attacks on WhatsApp and Signal highlight qualitatively different kinds of injection vulnerabilities, which yields enhanced breadth to our investigation of end-to-end encrypted backups. First, Signal uses a drastically different approach to serializing an SQLite database compared to WhatsApp, which leads to much less leakage and cross-user interactions (as far as our analyses have shown). Thus, there exists already in practice a diversity of encoding approaches, each with different levels of leakage. At the same time, WhatsApp uses a different encryption strategy than Signal: the latter does not monolithically encrypt the encoded database, but rather piecemeal. This results in leaking more granular information in the form of the plaintext length headers about the application state, which our injection attacks against Signal~exploit.

\paragraph{Mitigations} The most immediate mitigations for the attacks from
Sections~\ref{sec:dedup} and~\ref{sec:compression-attacks} would be to turn off
deduplication and compression.  Even though these would disable both leakage
channels completely, and thus shut the door for future attacks in these
contexts, the decrease in performance may be prohibitive. For example, turning
off both resulted in an 18x increase in the size of backups in our
testing environment. 

Another approach would be to avoid cross-user deduplication and compression, for
example by only compressing data associated within a conversation. Because
adversarially injected messages would only be deduplicated or compressed with
other information already available to the adversary, attacks would be
prevented. Whether this provides sufficient space savings for deployment is
unclear. 

A third approach would be to use padding in an attempt to make attacks harder.
For example, padding filenames to the maximum of 255 before encryption would
prevent our deduplication attack for dictionaries whose files are all the
same size. Similarly, one might try to add some amount of padding to file
contents to render their encrypted sizes uninformative to adversaries,
or add padding to a zlib-compressed
representation of the full database.  But padding approaches
are unlikely to prevent all cross-user leakage, and in other contexts such as network
traffic fingerprinting (see, e.g.,~\cite{dyer2012peek}) new attacks often broke
padding scheme recommendations. Thus, one would need some framework for carefully
reasoning about security. 

For deduplication in storage settings, Harnik et al.~\cite{harnik2010side}
suggested deduplicating only when the number of copies of a file reaches some
randomly assigned threshold. This might help mitigate attacks in our context
as well.

Turning to our attack exploiting FTS4 in WhatsApp, one could mitigate it by
not storing the index in the backup at all, and instead reconstructing it from the backed up messages. 

Finally, for the Signal attack, a simple mitigation is to encrypt the length
headers in a boundary-hiding way~\cite{boldyreva2012security}. This would result
in backups only leaking the total size of the backup, severely limiting what can be inferred.
This can be done using standard encryption mechanisms whose ciphertexts are indistinguishable 
from random bits (such as AES-GCM). This has now been adopted by Signal as a result 
of our disclosure, as we discuss below.

\paragraph{Disclosures} 
We responsibly disclosed our findings to both vendors, offering to discuss
countermeasures and work with them on timing of public disclosure. 
Signal acknowledged our vulnerability and have already
included hiding boundaries between ciphertexts in their v1 revision to their
Android backup file format.
WhatsApp acknowledged our vulnerabilities, but have not yet disclosed mitigations
plans.

\paragraph{Future work}
Our results add to a growing body of recent
work~\cite{patersonthree,albrecht2022practically,backendal2023mega,hogan2023dbreach}
to suggest that the community needs more attention on, and new approaches for, more holistically
analyzing the security of E2E encrypted applications. We note that this seems
particularly relevant as more such applications are being developed
and deployed, beyond E2E encrypted messaging. Examples include web-based productivity
tools like Google Docs and spreadsheets~\cite{googlecse,googlenewe2e}, Apple
iCloud encrypted backups (which also back up application
state)~\cite{icloudsecurity}, and more. As more client-side encryption arises in
applications, narrow investigation of individual protocols is insufficient to
understand the overall security of an application, and we will need new ways of
sussing out vulnerability to injection (and other) attacks, as well as new approaches to
guide the construction of applications in a way that achieves high assurance against them.

\section{Conclusion}
\label{sec:conclusion}
In this paper, we showed new injection attacks against E2E encrypted applications. We explored these attacks in the context of backups for
E2E encrypted messaging applications;
however, our injection threat model can be extended to E2E encrypted applications in general. 
We performed different experiments to demonstrate our attacks work as a proof-of-concept. The attacks
have some limitations, most notably they are rate-limited by the frequency of
backups and in some cases, their efficacy degrades in the presence of
noise from other
messages received by a target. Nevertheless, these attacks show that 
the desired level of E2E encryption security is not currently being met. 
We therefore believe
more work is needed both on ways to find injection vulnerabilities and
principled approaches that mitigate or, better yet, remove them entirely.

\bibliographystyle{plain}
\bibliography{reference}

\appendices

\section{Experimental Setup}\label{sec:a-methodology}
Our experimental setup had two parts: real clients, and a local simulation of the core client-side operations related to end-to-end encrypted backups. Since WhatsApp is not open source, the latter involved careful forensic analysis of their clients, which we describe below.

\subsection{Real Clients}\label{sec:a-experiments-real}
We used Android Emulator~\cite{androidemulator} to set up emulated Android phones running Android version 12, on an Apple laptop with an M2 chip running macOS Monterrey. Each phone had an associated (Google Voice) phone number and Google Drive account. We installed WhatsApp version 23.2.75 and Signal v6.22.8.

\paragraph{Signal}
We activated the Signal backup option, which stores backups locally. Then, we manually triggered the backups and exported them into Google Drive to inspect them outside the emulators. To better understand the structure of the backups, we decrypted them with an open-source decryptor~\cite{githubsignaldecryptor} and reviewed Signal's open-source implementation~\cite{githubsignalandroid}.

\paragraph{WhatsApp}
We extracted WhatsApp databases by downloading the backups from Google Drive with~\cite{githubwhatsappcloudextractor} and decrypting them with~\cite{githubwhatsappdecryptor}. By inspecting the format of the plaintext, we were able to determine the usage of zlib compression, which was evident from the magic header at the top of the file. Lastly, we decompressed this plaintext to extract the actual database file, which revealed their serialization method, SQLite parameters, schema, etc. With this pipeline, we could implement our attacks on real WhatsApp clients. 

This setup also let us inspect the state of the database very closely, and see exactly how different inputs change the internal database (e.g., how many bytes are added after an operation), which was crucial to implement some of our attacks. In addition, it let us collect real WhatsApp databases, under various setups, to test our local attacks appropriately.

\subsection{Local Simulation}\label{sec:a-experiments-local}
To set up our local simulation for WhatsApp, we reimplemented the core steps involved when sending a message and backing up data. For this, we compared many pairs of decrypted snapshots of the database, before an after each operation, and identified all differences between both.

To mock the internal changes after receiving a message, we sent various messages from one client to another, collecting a decrypted snapshot of the database before and after each (as described above). Then, we used sqldiff~\cite{sqldiff} to extract a list of SQL statements that transforms one database to the other, i.e., all state changes across all tables. Then, we manually inspected each individual change to understand what it entailed, and reimplemented this logic in a local API that runs the same list of SQLite commands with the appropriate payloads. Indeed, sending a message with our local API yield an analogous state to that of sending a message from a real WhatsApp client to another.

In addition, we empirically reverse-engineered the steps of the encryption process by closely inspecting and comparing multiple ciphertexts. We sanity-checked these findings using the (unmodified) decryptor tool~\cite{githubwhatsappdecryptor}: we downloaded and decrypted a database using our real pipeline, (re-)encrypted it in our local setup, and (re-)decrypted it with the same tool. This process was successful, and returned the same starting database. Our local API and encryption pipeline are thus analogous to sending a message and backing up the data in a real WhatsApp client, which means that our local experiments are an accurate representation of reality.

\end{document}